\documentclass{appolb}
\usepackage{graphicx}
\usepackage{amssymb}

\begin{document}
\eqsec  
\title{Generalized binomial transform applied to the divergent series%
}
\author{Hirofumi Yamada
\address{Division of Mathematics and Science, Chiba Institute of Technology, 
\\Shibazono 2-1-1, Narashino, Chiba 275-0023, Japan}
\\
}
\maketitle
\begin{abstract}
The divergent series for a function defined through Lapalce integral and the ground state energy of the quartic anharmonic oscillator to large orders are studied to test the generalized binomial transform which is the renamed version of $\delta$-expansion proposed recently.  We show that, by the use of the generalized binomial transform, the values of functions in the limit of zero of an argument is approximately computable from the series expansion around the infinity of the same argument.  In the Laplace integral, we investigate the subject in detail with the aid of Mellin transform.  In the anharmonic oscillator, we compute the strong coupling limit of the ground state energy an the expansion coefficients at strong coupling from the weak coupling perturbation series.  The obtained result is compared with that of the linear delta expansion.
\end{abstract}

\date{\today}

\section{Introduction}
The $\delta$-expansion proposed in ref. \cite{yam} has been considered so far on the discretized back ground.  In all applications of the method, the expansion in terms of the basic parameter has a finite radius of convergence such as the strong coupling expansion in the field theoretic models on lattice and high temperature expansion in magnetic systems \cite{yam2,yam3}.  The existence of the non-zero convergence radius plays an important role in the application of the $\delta$-expansion.  The aim of this paper is to investigate whether the method is effective in asymptotic series appearing in perturbation expansion.  Here we will apply the method to two models, a mathematical function defined through Laplace integral and the quantum mechanical anharmonic oscillator in which we focus on the computation of the ground state energy at strong coupling from the weak coupling perturbation theory to large perturbative orders. 

Before the argument, to avoid possible confusion, we like to rename the "$\delta$-expansion" used in \cite{yam} to the "generalized binomial transform" for the following reasons:   The anharmonic oscillator can be viewed as the Euclidean $1$-dimensional $\phi^4$ field theory.  The Hamiltonian then reads
\begin{equation}
H=\frac{1}{2}\Big(\frac{\partial \phi}{\partial q}\Big)^2+\frac{m^2}{2}\phi^2+\lambda \phi^4,
\label{hamiltonian}
\end{equation}
where we denoted the Euclidean time coordinate by $q$.  

The non-linear interaction is controlled by the coupling constant $\lambda$ and the un-perturbed current mass is given by $m$.  
The perturbation theory provides an expansion of any physical quantity in $\lambda$.  Due to the mass dimension $3$ carried by the coupling constant, expansion in $\lambda$ is actually the expansion in dimensionless parameter $\lambda/m^3$.  Thus it is apparent that the perturbative expansion is almost equivalent with the inverse-mass expansion.  There exists a novel computational technique called "linear delta ($\delta$) expansion", "optimized perturbation theory" or "variational perturbation theory" \cite{knp}.  Also the "order dependent mapping" \cite{sz} method, which includes "linear delta ($\delta$) expansion" in a specific fixing of the mapping, shares a similar feature.   In anharmonic oscillator, the "$\delta$-expansion" proposed in \cite{yam} has similarities to these techniques.  The conventional linear delta expansion introduces $\delta$ as the interpolating parameter by the substitutions $m^2\to m^2(1-\delta)$ and $\lambda\to \lambda\delta$.  The Hamiltonian to start with is
\begin{equation}
H(\delta)=\frac{1}{2}\Big(\frac{\partial \phi}{\partial q}\Big)^2+\frac{m^2}{2}\phi^2+\delta(-\frac{m^2}{2}\phi^2+\lambda \phi^4).
\label{lde-hamiltonian}
\end{equation}
Notice that the system at $\delta=0$ reduces to the free massive oscillator and the system at $\delta=1$ to the massless anharmonic oscillator (pure anharmonic oscillator).   One then regards $H(0)=\frac{1}{2}(\partial \phi/\partial q)^2+(m^2/2)\phi^2$ as the unperturbed part and expands the perturbation $\delta[-(m^2/2)\phi^2+\lambda \phi^4]$ as the power series in $\delta$.   The result is understood as the perturbative one of the interpolated system with the mass $m^2(1-\delta)$ and the coupling constant $\delta\lambda$.  Then, setting $\delta=1$, one may obtain nontrivial and effective estimates of physical observables in the massless limit (or the strong coupling limit).  The literatures on the method are quite many and see, for example, the paper \cite{knp} and references therein.  For the application of linear delta expansion on the lattice, see \cite{lattice}.

The "$\delta$-expansion" proposed in ref. \cite{yam} has been derived in the similar technique.  Suppose that $f(m^2)$ be given as the truncated series in $1/m^2$ to order $N$,
\begin{equation}
f_{N}(m^2)=\sum_{n=0}^{N}a_{n}\Big(\frac{1}{m^2}\Big)^n.
\label{series}
\end{equation}
By the re-scaling of the argument $m^2=(1-\delta)/t$, expanding $f_{N}((1-\delta)/t)$ in $\delta$ to the relevant order and setting $\delta=1$, one obtains the $\delta$-expansion of $f$ as the function of $t$ (See (\ref{binomial})).  This technique is first used on the lattice as a tool of dilatation of the continuum scaling region, where the argument is basically related to the lattice spacing.  On the other hand, the linear delta-expansion  is stemmed from the interpolation of two different systems.  Though the two tools share similar features and sometimes produce same results, they differ in the underlying concept and specific details, in particular when applied to physical systems on the lattice.  This is the main reason of renaming the "$\delta$-expansion" to "generalized binomial transform" or simply in short "binomial transform".

It would be in order to review briefly the generalized binomial transform:    In typical cases, the generalized binomial transform acts on the simple truncated power series (\ref{series}) as
\begin{equation}
{\cal B}_{N}[f_{N}(m^2)]=\sum_{n=0}^{N}a_{n}{N \choose n}t^n.
\label{binomial}
\end{equation}
That is, the coefficient $a_{n}$ in original series is multiplied by the binomial factor 
\begin{equation}
{N \choose n}=\frac{N!}{n!(N-n)!},\quad (0\le n\le N).
\label{binomial1}
\end{equation}
Thus, the binomial transform denoted by ${\cal B}_{N}$ is dependent on the perturbative order $N$.  For the sake of notational simplicity, we also use "bar" to imply the transform,
\begin{equation}
{\cal B}_{N}[f_{N}(m^2)]:=\bar f_{N}(t).
\end{equation}

At first sight, one may feel difficulty in understanding the effectivity of $\bar f_{N}(t)$ in extracting the quantities, the limit $\lim_{m\to 0}f(m^2)=f(0)$ or the critical exponent when divergent in power-like manner as $m\to 0$.  Though $\bar f_{N}(t)$ is just a polynomial in $t$, we found in some physics models that the asymptotic behavior of $f(m^2)$ as $m\to 0$ is observable in $\bar f_{N}(t)$ at {\it non-large $t$ region} and the marking quantities (the limit and critical exponents) can be estimated.  For instance in the square Ising model at temperature $1/\beta$, $\beta$ is expressed as a function of the mass-like parameter $M$ which is roughly the inverse of the square of the correlation length ($M$ is composed by the magnetic susceptibility $\chi$ and the second moment $\mu_{2}$ as $M=4\chi/\mu_{2}$).  The effective region of ${\cal B}_{N}[\beta(M)]=\bar\beta_{N}(t)$ in the $N\to\infty$ is numerically assured to be $0<t<0.25$ \cite{yam2} over which the function exhibits an extremely flat plateau and the inverse critical temperature $\beta_{c}$ is indicated at the stationary or almost stationary point of the function $\bar\beta_{N}(t)$.   For convergent series, it is implied that the limit $\lim_{N\to \infty}{\cal B}_{N}[f_{N}]$ is constant over $0<t<t_{c}$ for a certain $t_{c}$ and the function in the $N\to \infty$ limit has the shape like the step function with the finite range $(0,t_{c})$.    

Turning to the anharmonic oscillator, we deal with the perturbative expansion in powers of the coupling constant $\lambda$.   Then, remembering that the ground state energy $E(m,\lambda)$ has an asymptotic expansion in $\lambda/m^3$ with alternate signs, there are two crucial differences compared to the Ising case and 2D large $N$ vector model discussed in refs. \cite{yam,yam2,yam3}.  The first is that the expansion parameter has fractional powers of $1/m^2$ such as $(1/m^2)^{-1/2}$, $(1/m^2)^{5/2}$ and so on.  To handle these terms, we use the generalized binomial factor,
\begin{equation}
{N \choose s}=\frac{\Gamma(N+1)}{\Gamma(s+1)\Gamma(N-s+1)},
\label{binomial2}
\end{equation}  
where $s$ is real or complex when the extension is convenient or necessary, and the transformation rule is given by
\begin{equation}
{\cal B}_{N}[M^{-s}]={N \choose s}t^s,\quad M=m^2.
\label{binomial3}
\end{equation}
As we shall see in the next section, this rule is suitable when the function $f(M)$ of interest allows Mellin transform representation.

Second, as aforementioned, the original series has zero convergence radius and it is unclear at all whether the generalized binomial transform effectively works as before.   In this paper, we will demonstrate that a careful use of the principle of minimum sensitivity (PMS) \cite{stev} provides an accurate sequence of estimates to large enough orders.

This paper is organized as follows:  In the second section, we apply the binomial transform to a mathematical function which allows divergent expansion such that $(1/M)-2!(1/M)^2+3!(1/M)^3-\cdots$ and investigate in detail the computation of the limit $M\to 0$ from the divergent series expanded at $M=\infty$, the opposite end point of the argument contained in the range $(0,\infty)$.  The technique of Mellin transform representation is introduced to make the analysis transparent.  From this example, one can gain concrete feeling of how binomial transform works.  In the third section, we investigate the application of the binomial transform to the anharmonic oscillator.  We first review weak coupling perturbation expansion and consider its transform.  The linear delta-expansion is also mentioned and the difference is explained.  Explicit estimation will be worked out to the order $N=300$.  The sequence of the estimates indicates the strong evidence of the convergence to the most precise value to date, even though the region "effective" in the estimation shrinks as the order grows.  Also presented is the computation of the strong coupling coefficients and the result is compared with that from the conventional linear delta expansion.    The binomial transform related to the dilatation of the region around $\lambda=\infty$ is finally investigated.  The last section is devoted to the concluding remarks.

\section{A Laplace integral}
\subsection{Mellin transform}
In the use of transformation rule (\ref{binomial3}) to the closed form of function, Mellin transform plays an important role.  Given a function $f$ as the argument $M$, the representation through Mellin transform reads
\begin{equation}
f(M)=\int_{c-i\infty}^{c+i\infty}\frac{ds}{2\pi i}M^{-s}\varphi(s),
\label{mellin}
\end{equation}
where 
\begin{equation}
\varphi(s)=\int_{0}^{\infty}dM\, M^{s-1}f(M).
\end{equation}
In (\ref{mellin}), the integration contour in the complex $s$-plane is taken as the vertical one passing through $c\in \mathbb{R}$ and it is assumed that the integral (\ref{mellin}) exists in a certain vertical strip including the point $(c,0)$.  

The expansion of $f(M)$ at small  $M$ is given by the deformation of the contour to the left half-plane, by which residues at supposed poles leave the required series.  As well, the expansion in $1/M$ is obtained by the deformation of the contour to the right half-plane. 

When the Mellin transform representation is available, the generalized binomial transform is easy to implement.  We find the result from (\ref{binomial3}) that
\begin{eqnarray}
\bar f(t)&=&\int_{c-i\infty}^{c+i\infty}\frac{ds}{2\pi i}{\cal B}_{N}[M^{-s}]\varphi(s)\nonumber\\
&=&\int_{c-i\infty}^{c+i\infty}\frac{ds}{2\pi i}\frac{\Gamma(N+1)}{\Gamma(s+1)\Gamma(N-s+1)}t^{s}\varphi(s).
\end{eqnarray}
The kernel changes from $\varphi(s)$ to $\frac{\Gamma(N+1)}{\Gamma(s+1)\Gamma(N-s+1)}\varphi(s)$.   Deformation of the contour to the left half-plane gives the expansion of $\bar f(t)$ in $1/t$.  Thus, the large $M$ behavior of $f(M)$ corresponds to the small $t$ behavior of $\bar f(t)$.  If $\varphi(s)$ has single pole at $s=-L$ for positive integer $L$, expansion at small $M$ has the term $M^{L}$.  In contrast for $\bar f(t)$, the corresponding $t^{-L}$ term is absent since then the singularity is cancelled by $1/\Gamma(s+1)$.  If $\varphi(s)$ has double poles at $s=-L$, then there appears $M^{L}\log M$ in $f(M)$, but for $\bar f(t)$ just a power-like term $t^{L}$ remains and the associated residue becomes
\begin{equation}
\frac{(-1)^{L+1}\Gamma(N+1)\Gamma(L)}{\Gamma(N+L+1)}\varphi_{-2},
\label{reduction1}
\end{equation}
where the expansion $\varphi(s)=\varphi_{-2}/(s+L)^2+\cdots$ is supposed.  
It is crucial to note that the residue tends to vanish as $N^{-L}\varphi_{-2}$ as $N\to \infty$.  Surviving term is the residue at $s=0$ only, provided the pole is surrounded in the contour deformation.

\subsection{Generalized binomial transform applied to divergent expansion}
Let us consider the mathematical function for $M>0$ defined through Laplace integral given by
\begin{equation}
f(M)=M\int_{0}^{\infty}\frac{\omega\,e^{-M\omega}}{1+\omega} d\omega.
\label{Lap}
\end{equation}

From the well known result of Mellin transform,
\begin{equation}
e^{-M\omega}=\int_{c-i\infty}^{c+i\infty}\frac{ds}{2\pi i}(M\omega)^{-s}\Gamma(s),\quad \Re[s]>0,
\end{equation}
we obtain the following representation,
\begin{equation}
f(M)=\int_{c-i\infty}^{c+i\infty}\frac{ds}{2\pi i}M^{1-s}\Gamma(s)\Gamma(s-1)\Gamma(2-s),
\label{mellin0}
\end{equation}
where $s$ must obey $1<\Re[s]<2$.  
The integrand has double poles at $s=1, 0,-1,-2,\cdots$ and single poles at $s=2,3, \cdots$.  By the deformation of the integration contour to the left or the right, one obtains the series expansion in $M$ or $1/M$, respectively.  Due to the double multiplicity of poles, the single power of the logarithm appears in expansion in $M$.  The result reads from the residue computation that 
\begin{equation}
f(M)=1+M(\log M+\gamma_{E})+O(M^2\log M),
\end{equation}
where $\gamma_{E}$ denotes Euler-Mascheroni constant.  
The appearance of logarithm indicates the fact that the origin $M=0$ is a branch point.  This is understood by circulating the integration contour on the complex $\omega$-plane.   One then finds that the origin is a branch point and a circulation around the origin creates $2\pi i M e^{M}$, proving $f(M)$ be a multi-valued function. 

We notice that $\lim_{M\to +0}f(M)=1$ is given by the residue at $s=1$ and the pole $s=1$ is the first pole one encounters in the contour deformation to the left.  
On the other hand, $1/M$ expansion reads
\begin{equation}
f(M)=\frac{1!}{M}-\frac{2!}{M^2}+\frac{3!}{M^3}-\cdots.
\label{divexp}
\end{equation}
This series is divergent and deriving the asymptotic behavior of $f(M)$ at small enough $M$ requires a special technique such as Bore transform.   We like to show that the binomial transform converts the $1/M$ series into the series from which $f(0)$ can be approximately computable, even when the series is truncated at a given finite order.

The operation of the binomial transform is straightforward.  We find from (\ref{binomial3}) and (\ref{mellin0}) that
\begin{equation}
\bar f(t)=N!\int_{c-i\infty}^{c+i\infty}\frac{ds}{2\pi i}\frac{\Gamma(s-1)\Gamma(2-s)}{\Gamma(N-s+2)}t^{s-1}.
\end{equation}
The double poles of $f(M)$ at $s=0,-1,-2,\cdots$ have turned into the single poles and the expansion around $t=\infty$ becomes an infinite series without log.   The poles at $s=N+2,N+3,\cdots$ have disappeared due to the appearance of $1/\Gamma(N-s+2)$.  Thus, the series expansion in $t$ becomes a polynomial to the order $N$.  Then
\begin{equation}
\bar f(t)=\sum_{k=1}^{N}{ N \choose k}k!(-t)^k+R_{N}(t),
\end{equation}
where the function $R_{N}(t)$ represents the contribution from the deformed upward contour crossing at the positive real axis at some point located to the right of the largest pole $s=N+1$.  In realistic physical application, one does not have complete information and suffices truncated series to the order $N$.  Thus, we neglect the residual contribution $R_{N}$ and keep only the polynomial denoted by $\bar f_{N}(t)$,
\begin{equation}
\bar f_{N}(t)=\sum_{k=1}^{N}{ N \choose k}k!(-t)^k=N!\sum_{k=1}^{N}\frac{(-t)^{k}}{(N-k)!}.
\label{ap}
\end{equation}

For large $t$, gathering all residues of the poles $s=1, 0,-1,-2,\cdots$, we obtain 
\begin{eqnarray}
\bar f(t)
&=&\sum_{k=0}^{\infty}\frac{(-1)^{k}}{(N+1)\cdots (N+k)t^k}\nonumber\\
&=&N!\sum_{k=0}^{\infty}\frac{(-1)^{k}}{(N+k)! t^k}.
\label{ap1}
\end{eqnarray}
This is the expansion around $t=\infty$ and manifests $\bar f(t)$ be an entire function in the complex $1/t$-plane.   The function $\bar f(t)$ is single valued with no cut.  It is a crucial point here that all the coefficients except the one for the  leading term tend to zero when $N\to \infty$.  That is, $\bar f(t)$ tends to a uniform function,
\begin{equation}
\lim_{N\to \infty}\bar f(t)=1,\quad 0<t<\infty. 
\label{conv}
\end{equation}

In the contour deformation, the pole one first encounters is $s=1$ and the residue equals to $1$.  This is the limit $\bar f(\infty)$.  The agreement of $f(0)$ and $\bar f(\infty)$ is not of accidental because the residues at $s=1$ are kept equal with each other by the generalized binomial transform (by which $\Gamma(N+1)/\{\Gamma(s)\Gamma(N-s+2)\}$, which is equal to  $1$ at $s=1$, is created in the integrand).  
The function $\bar f(t)$ can be written as
\begin{equation}
\bar f(t)=N!\bigg[(-t)^{N} e^{-1/t}+\sum_{k=1}^{N}\frac{(-t)^{k}}{(N-k)!}\bigg].
\label{ap2}
\end{equation}
One can see that the second part agrees with $\bar f_{N}(t)$.  The first term has the essential singularity at $t=0$ and this is seen only in the deformation of the contour to the left plane.  The term does not allow expansion in $t$ and corresponds to $R_{N}(t)$ and leads us to  understand that $\bar f(t)$ expressed in (\ref{ap1}) and (\ref{ap2}) provides the exact result of generalized binomial transform of $f(M)$ (Contribution from the infinitely remote half-circle in the left-half plane disappears).

\subsection{Reduction of corrections to the asymptotic scaling}
Now, the point is whether $\bar f_{N}(t)$ to a given order $N$ is useful or not to simulate the dominant or leading behavior of $\bar f(t)$ at large enough $t$.  This is where the physics problems frequently arise.  By the numerical study of $\bar f_{N}(t)$ we find that the transformed series shows the improved behavior compared to the original truncated series of $f(M)$.  But the improvement is not sufficient and the estimation of $f(0)=\bar f(\infty)$ is not so good even at higher orders.  This is because the effect that the corrections in $\bar f(t)=1-\frac{1}{(N+1)}t^{-1}+\cdots$ fades out with the order $N$ is cancelled out by the shrinking of the effective region of $\bar f_{N}(t)$..  To suppress the correction, $L$th order linear differential equation,
\begin{equation}
\prod_{i=1}^{L}[1+p_{i}^{-1}(d/d\log t)]\bar f(t)=\bar f(\infty)+O(t^{-(L+1)}),
\label{lde}
\end{equation}
 is effective to subtract the corrections.  
Here, $p_{i}$ denotes the exponent of $\bar f(t)$ expanded at large $t$ and $p_{i}=i$ ($i=1,2,3,\cdots$).  We notice that the explicit expansion of $\bar f(t)$ at large $t$ is not needed here.  Used knowledge is just that the expansion is in the positive integer powers of $1/t$.  

The left hand side of (\ref{lde}) has small correction to $\bar f(\infty)=1$ at large $t$ of order $O(t^{-(L+1)})$.   Also at small $t$, the correction is expected to be reduced, since at large enough $N$ the coefficient of $t^{-(L+1)}$ vanishes as $(1/N)^{L+1}$ (see (\ref{reduction1})).  This suggests that $\bar f(t)\sim 1$ from large  to small $t$ region when $N$ is large.  We therefore replace $\bar f(t)$ in (\ref{lde}) by $\bar f_{N}(t)$, which is effective for small $t$, provided that at some order or above $\bar f_{N}(t)$ may be a good simulation of $\bar f(t)$.  Let us then denote
\begin{equation}
\psi_{L}=\prod_{i=1}^{L}[1+p_{i}^{-1}(d/d\log t)]\bar f_{N}(t).
\end{equation}   
By the input of exact values of $p_{i}$, we can indeed obtain better behaviors:  See FIG. 1 where $\psi_{L}$ for $L=0,1,2,3$ are plotted at $N=20$.   There appeared a plateau which grows flatter as the parameter numbers are increased.   However, as the order $N$ increases the plateau becomes narrower and the center moves to the origin, which is the influence of the divergent nature of $1/M$ expansion.   The plateau represents, due to the successful elimination of corrections, the leading term in the $t\to \infty$, $\bar f(\infty)=1$.  It is  natural to estimate $\bar f(\infty)$ on the unique top on the plateau and this protocol is called the principle of minimum sensitivity (PMS) \cite{stev}.   The results of estimation using PMS are summarized in table 1.
\begin{figure}
\centering
\includegraphics[scale=0.85]{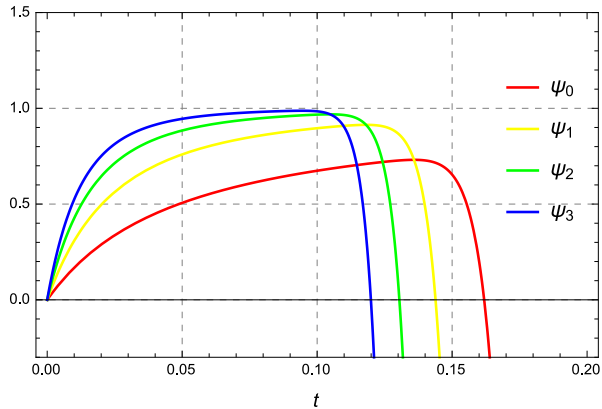}
\caption{$20$th order plots of $\psi_{L}(t)=\prod_{i=1}^{L}[1+p_{i}^{-1}(d/d\log t)]\bar f_{N}(t)$ ($L=0,1,2,3$) with the correct values $p_{1}=1$, $p_{2}=2$ and $p_{3}=3$.}
\end{figure}
\begin{table}
\caption{Estimation of $f(0)=\bar f(\infty)=1$ through $\psi_{L}(t)$ using principle of minimum sensitivity.  We performed computation to $300$th order, while the shown results are up to $40$th.   The last column labeled by "$\infty$ (extrapolated)" indicates the extrapolated value from the $290$th and $300$th results via the ansatz $\bar f(\infty)(1-f_{1}N^{-1})$.}
\begin{center}
\begin{tabular}{clll}
\hline\noalign{\smallskip}
$N$  &\quad $L=0$ &\quad $L=1$ &\quad $L=2$   \\
\noalign{\smallskip}\hline\noalign{\smallskip}
$10$   &\quad  0.69276626  &\quad 0.87951576  &\quad 0.94485674 \\
$20$   &\quad  0.73101017  &\quad 0.91367909   &\quad 0.96853507 \\
$30$   &\quad  0.74556188  &\quad 0.92559505 &\quad 0.97580002 \\
$40$   &\quad  0.75337822  &\quad 0.93172054 &\quad 0.97928804 \\
$\infty$ (extrapolated)   &\quad  0.78151  &\quad 0.95212  &\quad 0.98950 \\
\noalign{\smallskip}\hline
\noalign{\smallskip}\hline
$N$  &\quad $L=3$ &\quad $L=4$ &\quad $L=5$   \\
\noalign{\smallskip}\hline\noalign{\smallskip}
$10$   &\quad  0.97185783  &\quad 0.98441767  &\quad 0.99080598 \\
$20$   &\quad  0.98733751 &\quad 0.99448001  &\quad 0.99742866 \\
$30$   &\quad  0.99141838  &\quad 0.99672795 &\quad 0.99867246 \\
$40$   &\quad  0.99321942  &\quad 0.99763006 &\quad 0.99912290 \\
$\infty$ (extrapolated)   &\quad  0.99771  &\quad 0.99950  &\quad 0.99989 \\
\noalign{\smallskip}\hline
\end{tabular}
\end{center}
\end{table}
\begin{figure}
\centering
\includegraphics[scale=0.80]{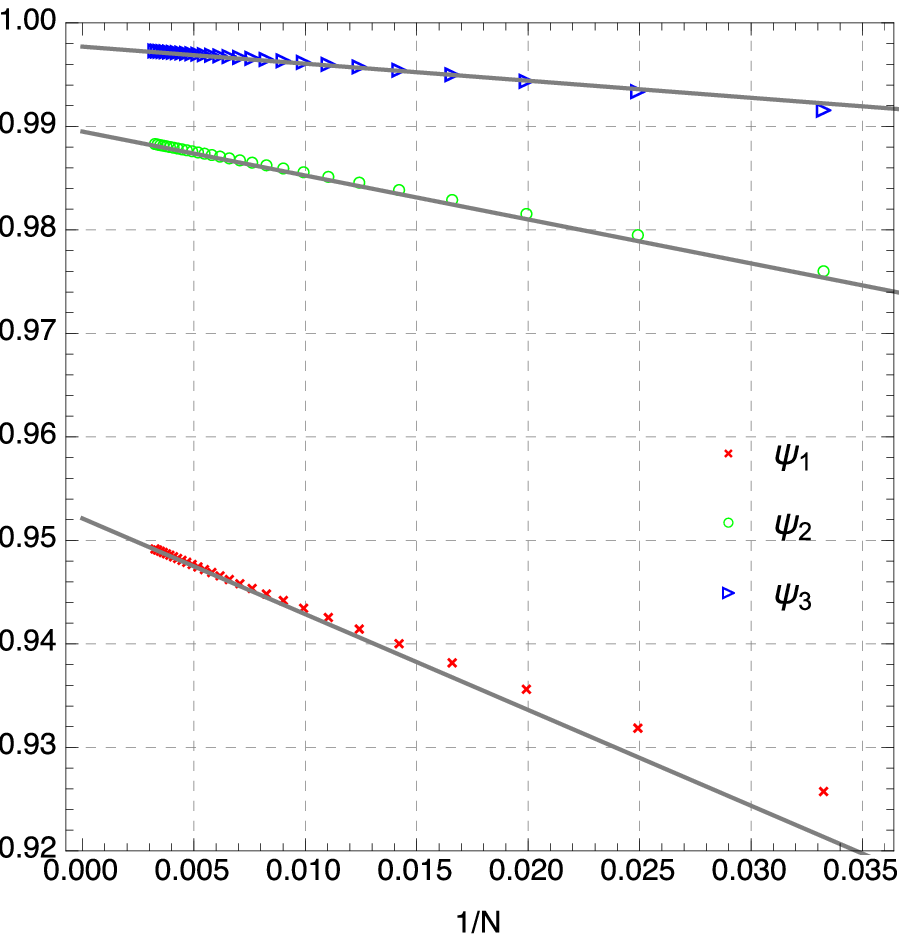}
\caption{Plots of estimation sequence v.s. $1/N$ for $L=1,2,3$.  The solid gray lines represent the fitted lines obtained via the ansatz $\bar f(\infty)(1-f_{1}N^{-1})$ from the results at $290$th and $300$th.}
\end{figure}

The sequence of $L\ge 1$ exhibits remarkable improvement over the plane $L=0$ sequence.    As many exponents are incorporated as the parameters of $\psi_{L}$, the accuracy becomes higher at all orders.  However, the convergence issue is subtle up to the $300$th oder which is the highest order estimation we have performed.   To settle the issue, we have done the fitting assumed the ansatz, $\bar f(\infty)(1-f_{1}N^{-1})$.  From the sample at $N=290$ and $300$, which are the highest computation orders, we obtained the extrapolated values listed in table 1.  In FIG. 2 we have shown the plots of the estimated sequence at $L=1,2,3$ and the obtained fitted lines.  It is confirmed that, at $L=0\sim 5$, though larger $L$ provides better and accurate approximation, the limits extrapolated do not agree with the exact value $\bar f(\infty)=1$.  The origin of this discrepancy is again the divergent nature of $f(M)$ in $1/M$ expansion, which reflects the narrowness and movement to the origin of the plateau and the top of the plateau fails to attain to the hight of $\bar f(\infty)=1$.    

The features described so far can be explained analytically as follows:  First let us consider the asymptotic behavior of the PMS solution $t^{*}$ in the $N\to \infty$ limit.  From numerical analysis, we find that $t^{*}$ decreases as $t^*\sim c_{L}/N$ with $c_{L}$ constant.  The values of $c_{L}$ are $c_{0}\sim 3.4855$, $c_{1}\sim 3.4252$, $c_{2}\sim 3.3698$ and so on.  Though these values are obtained at each $L$ with respective highest-order ($300$th) values of $t^{*}$, they are in fact dependent on $N$.  Since the behaviors of the sequences of $t^{*}\times N$ to the order $N=300$ are monotonically increasing with $N$ for all $L$ examined, the values of $c_{L}$ indicated in the $N\to \infty$ limit would be slightly larger than the above values.  We can actually infer the value of true $c_{L}$ to be identified only in the $N\to \infty$ limit from (\ref{ap2}).  For example consider the case $L=0$.  The residual part $R_{N}$ is given by
\begin{equation}
R_{N}=\bar f(t)-\bar f_{N}(t)=N!(-t)^N e^{-1/t}.
\end{equation}
Substituting the ansatz $t^{*}=c_{0}/N$ into above, we obtain $R_{N}=N!(-c_{0}/N)^Ne^{-N/c_{0}}$ and at large enough $N$, from Stirling's formula, $R_{N}\sim \sqrt{2\pi N}(-c_{0}/e^{(1+1/c_{0})})^N$.  Thus, if $R_{M}(t^*)\to 0$ as $N\to \infty$, the condition $c_{0}/e^{(1+1/c_{0})}<1$ is deduced.  The maximally allowed value of $c_{0}=c_{0,max}$ is then found as the solution of $\log c_{0,max}=1+1/c_{0,max}$, giving $c_{0,max}=3.591121476668622\cdots$.  For $L\ge 1$, the same analysis can be carried through and the result of upper limit of $c_{L}$ is found to be independent of $L$.  Thus, we conclude
\begin{equation}
c_{L, max}=3.591121476668622\cdots.
\label{limit-c}
\end{equation}
The values of $c_{L}$ for $L=1\sim 5$ obtained at $N=300$ are all under and close to the above limit.  Now as mentioned before, the estimated $c_{L}$ grows with the order and surely tends to the value very close or exact to $c_{L,max}$ at $L=0,1,2,\cdots 5$.  We hence assume that estimated $c_{L}$ converges to $c_{L,max}$ and compute the limit of the sequence of $\bar f^{*}$.  First of all, we note that $\bar f(t)$ can be used in this study instead of $\bar f_{N}$ itself since $R_{N}\to 0$ ($N\to \infty$) is assured.  Then, substituting $t=c_{L}/N$ into $\psi_{L}=N!\sum_{n=0}^{\infty}(-1)^n/\{(N+n)!t^n\}$ and expanding the result in $1/N$, we obtain
\begin{equation}
\psi_{L}(t^{*})=1-\frac{1}{(1+c_{L})^L}+O(1/N).
\end{equation}
Substitution of $c_{L,max}$ into $c_{L}$ produces
\begin{equation}
\lim_{N\to \infty}\psi_{L}(t^{*})=1-\frac{1}{(1+3.591121476668622\cdots)^L}.
\end{equation}
One finds that the above result agrees with the corresponding result indicated by "$\infty$(extrapolated)" in the last row in table 1.

The use of the exact values of the exponent is possible only when we know what values they are.  In the realistic physical situation in field theoretic and statistical models, the exponents are not exactly known, of course.  In this case, one approach is to resort to extended principle of minimum sensitivity, where the exponents are fixed as to make the higher order derivatives of $\psi_{L}$ be zero at the estimation point $t^{*}$ \cite{knp,yam2,yam3}.  In the present example, however, the approach fails.  It is because the higher order derivatives themselves do not reach enough scaling behaviors.

\subsection{Estimation via Pad\'e approximant}
As the second approach, we attempt another extrapolation scheme by Pad\'e method \cite{baker}.  Pad\'e approximants approximates a given function by the ratio of two polynomials in accordance with the truncated Taylor series. 

Here we utilize Pad\'e approximants constructed from the series $\bar f_{N}(t)$.  It should be remind here that for the estimation of $f(0)=\bar f(\infty)$, the best Pad\'e approximants among entries in Pad\'e table is the diagonal one at even $N$, since we can take the limit $t\to\infty$ and the result directly provides the estimation of $\bar f(\infty)=1$.  To define the protocol clearly, let us denote the Pad\'e approximant of $N=\rho+\tau$ decomposition as $\bar f_{N}[\rho/\tau]$.  Here the degree of the numerator polynomial is equal or less than $\rho$ and the degree of the denominator polynomial is equal or less than $\tau$.   The estimate via diagonal approximant with even $N$ is defined by
\begin{equation}
\bar f(\infty)=\lim_{t\to \infty}\bar f_{N}[\rho/\rho],\quad \rho=N/2.
\end{equation}
For example, at $N=10$,
\begin{equation}
\bar f_{N}[5/5]=\frac{10t+160t^2+1470 t^3+6960 t^4+15240 t^5}{1+25t+300t^2+2100 t^3+8400t^4+15120 t^5}
\end{equation}
and
\begin{equation}
\lim_{t\to \infty}\bar f_{N}[5/5]=\frac{127}{126}=1.0079365079\cdots.
\end{equation}

The same method is also used for $f_{N}(M)$, the original truncated series.  
 We estimated in the cases $(\rho,\tau)=(5,5),(10,10),(15,15),(20,20),(25,25)$ for both $f_{N}(M)$ and $\bar f_{N}(t)$.  The result is shown in table 2.  It is clearly seen that for $f_{N}(M)$ the sequence is monotonically increasing and shows tendency of approaching to $1$.  Actually, we find from numerical work that the estimate at $N$th order is given by $N/(N+2)=1-2/N+\cdots$.   The convergence speed is thus slow.   As for $\bar f_{N}(t)$, the convergence tendency is strongly exhibited and in particular the accuracy is excellent.  
\begin{figure}
\centering
\includegraphics[scale=0.85]{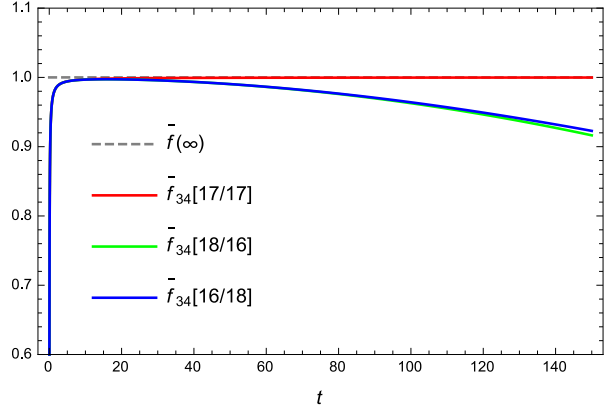}
\caption{Pad\'e apprpximants $\bar f_{34}[17/17]$, $\bar f_{34}[18/16]$ and $\bar f_{34}[16/18]$.}
\end{figure}
\begin{table}
\caption{Estimation of $f(0)=\bar f(\infty)=1$ using diagonal Pad\'e approximants of $f_{N}(M)$ and $\bar f_{N}(t)$.}
\begin{center}
\begin{tabular}{cll}
\hline\noalign{\smallskip}
$N$  &\quad $f_{N}(M)$ &\quad $\bar f_{N}(t)$   \\
\noalign{\smallskip}\hline\noalign{\smallskip}
$10$   &\quad  0.8333333  &\quad 1.0079365 \\
$20$   &\quad  0.9090909  &\quad 0.9999891749  \\
$30$   &\quad  0.9375000  &\quad 1.00000001289  \\
$40$   &\quad  0.9523809  &\quad 0.99999999998549  \\
$50$   &\quad  0.9615385  &\quad 1.0000000000000158  \\
\noalign{\smallskip}\hline
\end{tabular}
\end{center}
\end{table}
We note that the sequence here shows small oscillation with the minimum period.  At $N=2+4K$ ($K=0,1,2,\cdots$), the sequence approaches to $1$ from above and at $N=4+4K$ from below. In each sub-sequences, the error is exponentially small with the $N$ dependence roughly found to be $\log_{e}|f_{N}^{*}-1|\sim 3.4-0.693\times N$ for both subsequences ($f_{N}^{*}$ denotes the estimate at order $N$).  

The reliability of results through diagonal approximants becomes solid when the near diagonal ones, $\bar f_{N}[\rho/\tau]$ with $|\rho-\tau|=1$ or $2$, show broad plateaus.  We have observed from orders $N\sim 30$ or larger, $\bar f_{N}[\frac{N}{2}+1/\frac{N}{2}-1]$ and $\bar f_{N}[\frac{N}{2}-1/\frac{N}{2}+1]$ for even $N$ exhibit large plateaus.  See the diagonal and near-diagonal Pad\'e approximants in the plot (FIG. 3).  The reference values from the near diagonal approximants are obtained by the stationary values (local maximum in these cases) of $\bar f_{N}[\frac{N}{2}+1/\frac{N}{2}-1]$ and $\bar f_{N}[\frac{N}{2}-1/\frac{N}{2}+1]$.  At $N=34$, they are  $0.9973217\cdots$ (at $t=15.79656\cdots$) and $0.9973225\cdots$ (at $t=15.80567\cdots$), respectively.  These values are similar in accuracy to the estimates via $\psi_{4}$ case presented in the previous subsection.
\begin{figure}
\centering
\includegraphics[scale=0.80]{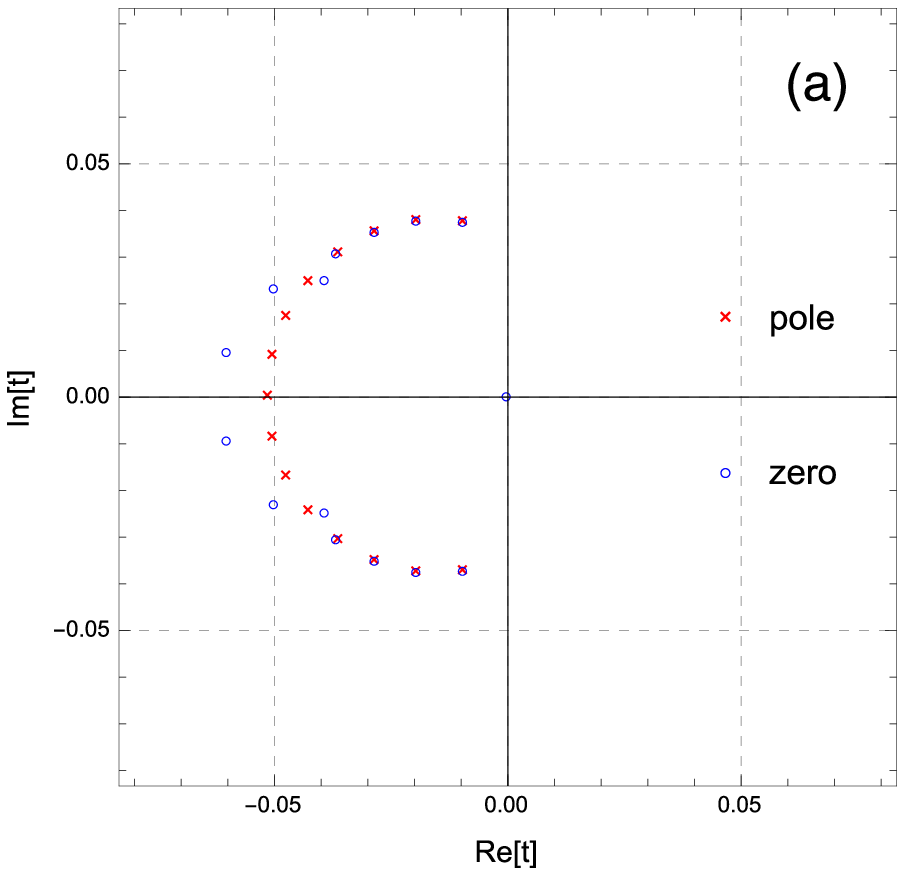}
\includegraphics[scale=0.80]{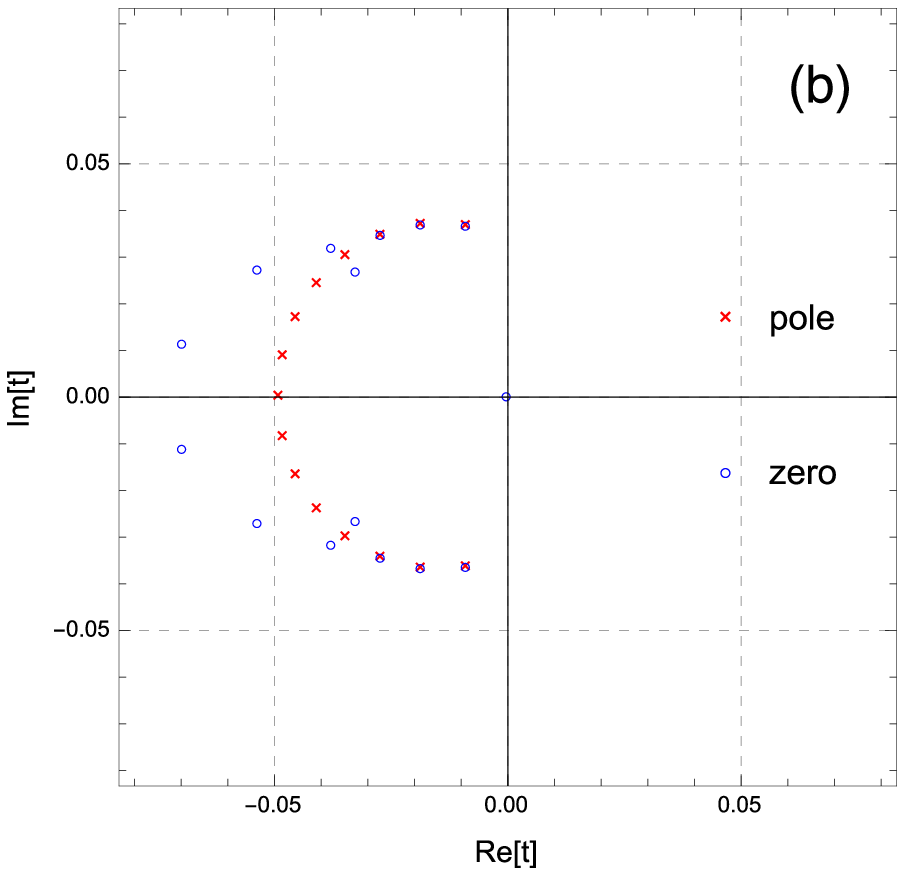}
\caption{Zeros and poles of Pad\'e apprpximants (a) $\psi_{0}[15/15]=\bar f_{30}[15/15]$ and (b) $\psi_{1}[15/15]$.}
\end{figure}

At first sight, one might think that diagonal Pad\'e approximants of $\psi_{L}=\prod_{i=1}^{L}[1+p_{i}^{-1}(d/d\log t)]\bar f_{N}(t)$ would be more suitable for $L=1,2,3,\cdots$.  Produced results are indeed accurate but not better than $\psi_{0}$.  The reason may be found by the enumeration of zeros and poles of the diagonal Pad\'e approximants.  The distribution of zeros and poles at $N=30$ are depicted in FIG. 4.  We see that for $\psi_{0}[15/15]=\bar f_{N}[15/15]$, the $8$ poles in the left half-plane are approximately cancelled by the zeros nearby them.  On the other hand for $\psi_{1}[15/15]=\{[1+p_{1}^{-1}(d/d\log t)]\bar f_{N}\}[15/15]$, the approximate cancellation occurs for $6$ pairs.  Since the existence of bare poles would affect the behavior of diagonal Pad\'e approximants on the positive real axis, it is better when the number of bare poles are small and located in remote place (the locations of poles of $\psi_{1}[15/15]$ are slightly inside of those of $\psi_{0}[15/15]$).  We can thus roughly understand why $\psi_{0}[15/15]$ provides better estimation.  Pad\'e approximants of the same construction for $f_{N}(M)$, $\prod_{i=1}^{L}[1-p_{i}^{-1}(d/d\log M)]f_{N}(M)$, prove improvement for larger $L$.  Actually, we have obtained analytic results of $N$th order estimate inferred from numerical study that $N(N+6)/\{(N+2)(N+4)\}$ and $N(N^2+12N+44)/\{(N+2)(N+4)(N+6)\}$ for $L=1$ and $2$, respectively.  The correction to $1$ is respectively $O(1/N^2)$ and $O(1/N^3)$.

We now conclude that diagonal Pad\'e approximants of $\bar f_{N}$ yield accurate convergent sequence of estimate.

\section{Anharmonic oscillator}
\subsection{Perturbative expansion of the ground state energy}
The perturbative ground state energy $E(m,\lambda)$ is given in the form
\begin{equation}
E(m,\lambda)=m\sum_{n=0}^{\infty}a_{n}\Big(\frac{\lambda}{m^3}\Big)^n.
\label{perturbation}
\label{energy}
\end{equation}
The coefficient $a_{n}$ can be computed from the recursion technique due to Bender and Wu \cite{bw}.  For instance, the first several coefficients read
\begin{equation}
a_{0}=\frac{1}{2}, \quad a_{1}=\frac{3}{4},\quad a_{2}=-\frac{21}{8},\quad a_{3}=\frac{333}{16}.
\end{equation}
We have generated the first $300$ coefficients exactly and use the result in the following studies.

It was shown in ref. \cite{bw} that the coefficient grows with the order $n$ as
\begin{equation}
a_{n}\sim -\frac{\sqrt{6}}{\pi^{3/2}}(-3)^{n}\Gamma(n+1/2),
\end{equation}
indicating zero convergence radius with the alternate coefficients.  In this paper, we deal with the truncated series to $N$th order perturbative expansion written as
\begin{equation}
E_{N}(m,\lambda)=m\sum_{n=0}^{N}a_{n}\Big(\frac{\lambda}{m^3}\Big)^n.
\end{equation}
The perturbative truncation order means the number of included terms and it is matched to the parameter $N$ involved in the generalized binomial factor (\ref{binomial2}).

\subsection{Binomial transform and linear delta expansion}
In accordance with the dilatation by the re-scaling of the square of the mass $m^2=(1-\delta)/t$, we describe the perturbative series in terms of $x$ defined by
\begin{equation}
x=\frac{\lambda^{2/3}}{m^2}.
\end{equation}
The perturbative expansion is not a simple series expansion with positive integer powers but a singular expansion with  
fractional powers such as $x^{(3n-1)/2}$ $(n=0,1,2,\cdots)$.  With respect to such a singular series, 
the binomial transform is defined with (\ref{binomial2}) as
\begin{equation}
E_{N}(x,\lambda)=\lambda^{1/3}\sum_{n=0}^{N}a_{n}x^{(3n-1)/2}\to \lambda^{1/3}\sum_{n=0}^{N}\bar a_{n} t^{(3n-1)/2},
\end{equation}
where the coefficient $\bar a_{n}$ is given by
\begin{eqnarray}
\bar a_{n}&=&a_{n}{N \choose \frac{3n-1}{2}}\nonumber\\
&=&a_{n}\frac{\Gamma(N+1)}{\Gamma(\frac{3n-1}{2}+1)\Gamma(N-\frac{3n-1}{2}+1)}.
\end{eqnarray}
That is, we obtain
\begin{equation}
\bar E_{N}(t)=\lambda^{1/3}\sum_{n=0}^{N}\bar a_{n} t^{(3n-1)/2}.
\label{energy2}
\end{equation}

The generalized binomial transform possesses a few characteristic features which differ from the linear delta expansion as below:  The first is that the factor $1/\Gamma(N-\frac{3n-1}{2}+1)$ becomes zero for some values of $N$ $(\ge 3)$ and $n$.  It vanishes for $(N, n)=(3,3), (5,5), (6,5), (7,7), (8,7), (9,7), (9,9)$ and so on.  This leads that a subset of terms in the original expansion is eliminated.   Second, the factor takes negative values for various sets of $(N, n)$ such as $(N,n)=(4,4), (6,6), (7,6), (8,8), (9,8)$ and so on.  The negative binomial factor changes the sign of the coefficients and rigorous alternativeness is slightly broken.  The original series is modified in this manner.  

On the contrast, the linear delta-expansion does not change the sign.  Some explanation would be needed here:  Let us denote the result of linear delta-expansion be
\begin{equation}
E_{LDE, N}(m)=\lambda^{1/3}\sum_{n=1}^{N}a_{n}C_{N,n}(\lambda^{2/3}/m^2)^{(3n-1)/2}.
\label{linear delta expansion}
\end{equation}
Here remind that the factor $C_{N,n}$ representing the modification comes from the term $m(\lambda/m^3)^n$ through the linear delta-expansion.  One can obtain $C_{N,n}$ from the plain perturbative series (\ref{perturbation}) by the shifts $\lambda\to \lambda\delta$ and $m^2\to m^2(1-\delta)$.  Then, $m(\lambda/m^3)^n\to m(\lambda/m^3)^n\delta^n(1-\delta)^{-(3n-1)/2}$.   The expansion of $\delta^{n}(1-\delta)^{-(3n-1)/2}$ in $\delta$ to the order $N$ and setting $\delta=1$ gives $C_{N,n}$.   For example at $n=0$, we shall expand such that $(1-\delta)^{1/2}=1-\frac{1}{2}\delta-\sum_{k=1}^{N}\frac{(2k-1)!}{2^{2k-1}k!(k-1)!}\delta^k$.  Then, putting $\delta=1$, the summation of the resulting series gives $C_{N,0}=(2N)!/\{2^{2N}(N!)^2\}$.  For general $N$, $C_{N,n}$ is obtained explicitly as \cite{bgn}
\begin{equation}
C_{N,n}={N+\frac{n-1}{2} \choose \frac{3n-1}{2}}=\frac{\Gamma(N+\frac{n+1}{2})}{\Gamma(\frac{3n+1}{2})\Gamma(N-n+1)}.
\end{equation}
 As may be clear from the above procedure, the result ensures that $C_{N,n}>0$ at any finite order $N$ for all $n=0,1,2,\cdots,N$.  We note that the factor $C_{N,n}$ is a rational number.  On the other hand, ${N \choose (3n-1)/2}$ multiplied to the original expansion coefficient in the binomial transform includes $\pi$ for odd $n$.  
 
 For further quantitative comparison, we have plotted the ratio $R_{N,n}={N \choose (3n-1)/2}/C_{N,n}$ for $n=0,1,2$ and $3$ against $N$ in FIG. 5. 
\begin{figure}
\centering
\includegraphics[scale=0.85]{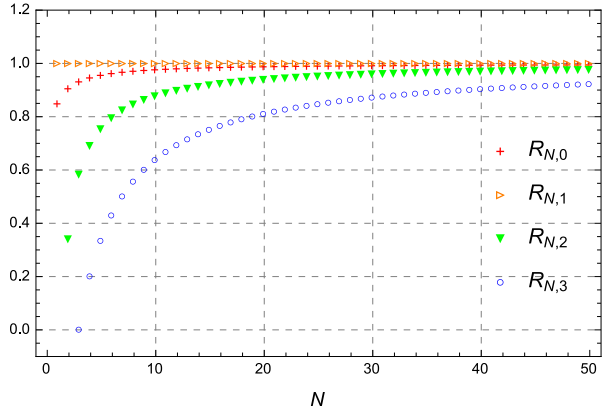}
\caption{Ratio plot of the leading and next-to-the leading order coefficients $R_{N,n}={N \choose (3n-1)/2}/C_{N,n}$ for $n=0,1,2,3$.}
\end{figure}
The ratio converges to unity in the $N\to \infty$ limit as $R_{N,n}=1-(n-1)(3n-1)/N+O(N^{-2})$, but the difference is not negligible at finite orders except for $n=1$ ($R_{N,1}=1$ to all orders).  

The convergence in the linear delta-expansion is proved in ref. \cite{gks}.  As for the generalized binomial transform method, the proof is not obtained.  However, large order numerical study provides convincing affirmative result on the convergence issue under PMS protocol by the comparison with the results of sequence in the linear delta-expansion.

\subsection{Computation of the ground state energy}
We now use $\bar E_{N}(t)$ to estimate the massless limit (or the strong coupling limit) of the ground state energy,
\begin{equation}
\lim_{m\to 0}E(m,\lambda)={\cal E}\lambda^{1/3},
\label{limit}
\end{equation} 
where ${\cal E}$ is given by Vinette and Cizek \cite{vin} to extreme accuracy \cite{wen},
\begin{eqnarray}
{\cal E}&=&0.66798625915577710827096201691986019943\nonumber\\
& &04049369840604559766608.
\label{energy3}
\end{eqnarray}

Before explicit computation, let us see how binomial transformed energy behaves against $t$.  FIG. 6 shows the plot of $\bar E_{N}(t)$ at $N=10,20$ and $30$.
\begin{figure}
\centering
\includegraphics[scale=0.85]{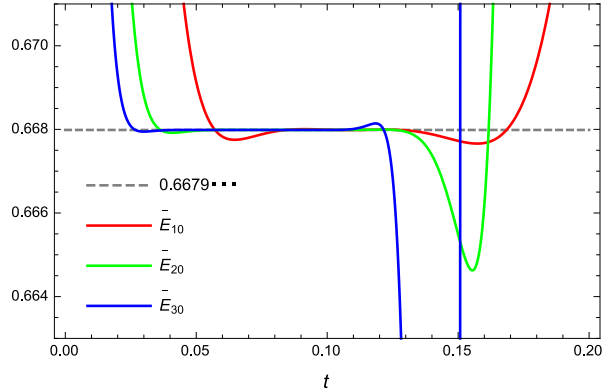}
\caption{Plot of $\bar E_{N}(t)$ with $\lambda=1$ at $N=10,20$ and $30$.}
\end{figure}
It is explicitly shown that $\bar E_{N}(t)$ clearly signals the correct value already at rather small order around $N=10$.  The value to be identified as the estimate of ${\cal E}$ in the plotted curves are implied by the plateaus.  The width of a plateau shrinks as the order grows and this feature reflects the asymptotic nature of the original perturbative series.
  
We notice then the problem pointed out by Neveu in \cite{neveu} that the plateau exhibits weak oscillation with tiny amplitudes.  The oscillation may be embarrassing indeed, since it leads to the non-uniqueness of the stationary solution under PMS protocol.   In ref. \cite{knp}, Kneur, Neveu and Pinto proposed an interesting prescription to terminate this oscillation by introducing additional parameters in the linear delta expansion.   Their idea is to generalize the simple prescription $m^2\to m^2(1-\delta)$ to the one involving more parameters such as $m^2\to m^2(1-\delta)(1+(a-1)\delta+\sum_{n=1}b_{n}\delta^{n+1})$ \cite{comment1}. For example, at the second order, modification is to use the shift $m^2\to m^2(1-\delta)(1+(a-1)\delta)$ and expand $\delta$ as in the conventional manner.  It is possible to adjust $a$ such that only single real-valued solution, the solution satisfying $(\partial/\partial m^2)E_{LDE,N}=(\partial/\partial m^2)^2 E_{LDE,N}=0$ exists.  At the third order, they found it suffice to use $m^2\to m^2(1-\delta)(1+(a-1)\delta+b_{1}\delta^2)$ and seek the unique solution obeying $(\partial/\partial m^2)E_{LDE,N}=(\partial/\partial m^2)^2 E_{LDE,N}=(\partial/\partial m^2)^3 E_{LDE,N}=0$ under the adjustment of $a$ and $b_{1}$.   The result was successful at low orders but turned out to getting worse at higher orders \cite{knp}.

We like to remark on this problem that, without introducing additional parameters, even many oscillations occur and many candidates appear, the best optimal estimation point can be detected by carefully observing the derivatives of $\bar E_{N}$;  See FIG. 7 where the first order derivative $\bar E^{(1)}_{N}=(\partial/\partial \log t)\bar E_{N}$ is plotted at $N=23$ and $50$.  Seeing the plot, we find that there exists a narrow region within the plateau that the first order derivative is oscillating with smallest amplitude.
\begin{figure}
\centering
\includegraphics[scale=0.85]{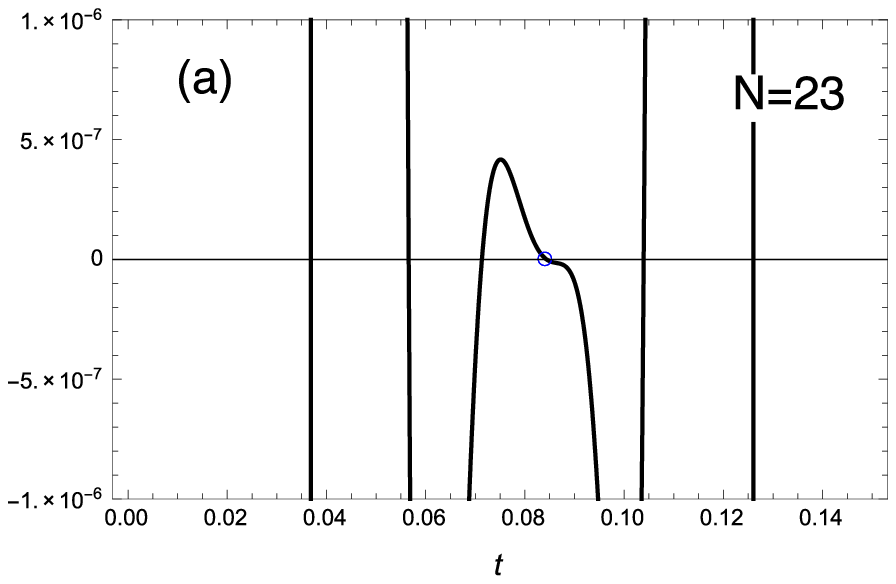}
\includegraphics[scale=0.85]{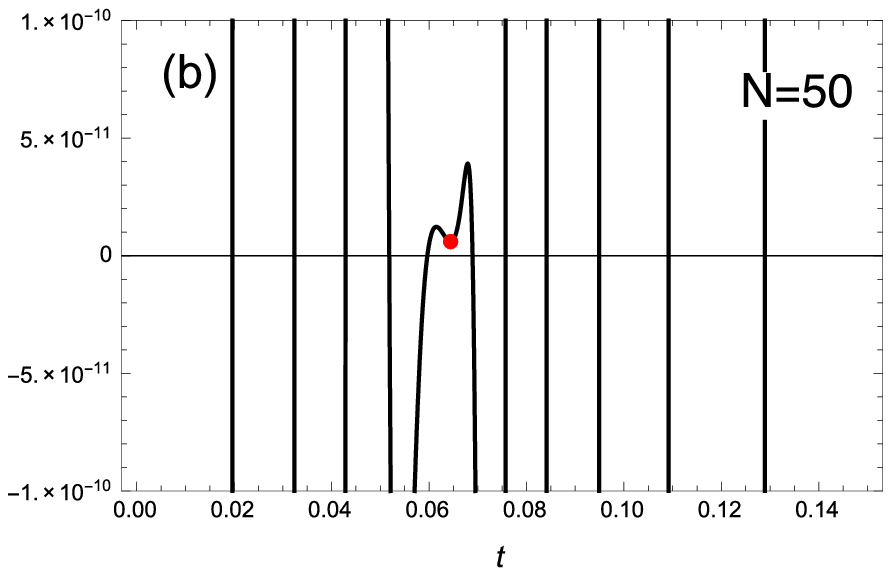}
\caption{First order derivatives $\bar E_{N}^{(1)}(t)$ at $N=23$ and $50$.  Within the plotted range, $\bar E_{23}^{(1)}$ has 6 solutions and $\bar E_{50}^{(1)}$ has $11$ solutions.   Optimal stationary or almost stationary points are indicated by the blue circle ($23$th) and filled red blob ($50$th) for the estimation of ${\cal E}$.    At $N=23$, the point is at $\bar E_{N}^{(1)}=0$ and $|\bar E_{N}^{(2)}|\ll 1$.  At $N=50$, the point is at $|\bar E_{N}^{(1)}|\ll 1$ and $\bar E_{N}^{(2)}=0$.  The later point corresponds to the reflection point with very small gradient.}
\end{figure}
With the increase of the order, the oscillatory wave becomes dense and a new oscillation wave  seems to be born from the region, as signaled by the smallest amplitude of the first derivative.  From this observation, we pose an assumption that the "center" of the set of zeros of $\bar E_{N}^{(1)}(t)$ be optimal as the estimation point.   In the case shown in FIG. 7(a), it is natural that the point indicated by the blue circle is optimal among other stationary points.  
In the case shown in FIG. 7(b) on the other hand, the point indicated by the red filled circle exhibits the tendency in next few orders that it goes down and across the horizontal axis, creating a new stationary point.  We therefore consider the red-marked point should be considered as the optimal estimation point, even though the first derivative is not zero at the point (Note that the value of the first derivative is extremely small there in magnitude).  

It is interesting to consider the complex extension of $\bar E_{N}(t)$ denoted as $\bar E_{N}(z)$ ($z\in  \mathbb{C}$) where $z=t^{3/2}$.  Numerically solving $\bar E_{N}^{(1)}(z)=0$ at $N=50$, we have plotted the solutions in the $z$-plane with blue circles in FIG. 8.  Red filled circles indicate the solutions of $\bar E_{N}^{(2)}(z)=[(\partial/\partial \log t)^2\bar E_{N}]_{t\to z}=0$.  Now, the point is that there exists a small area in which the arc-shaped sequence of complex zeros and the set of real zeros on the positive real axis are intersected. 
\begin{figure}
\centering
\includegraphics[scale=0.75]{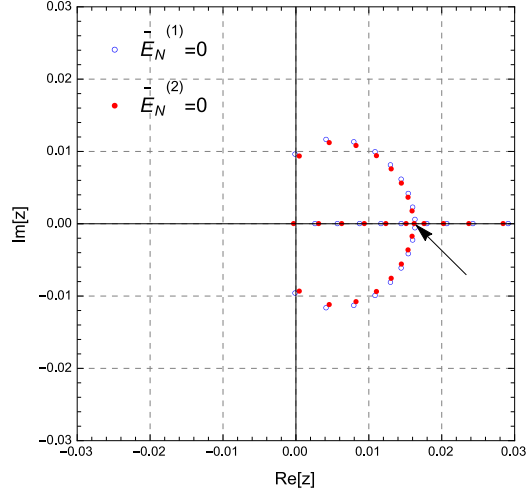}
\caption{The plot of zeros of the first and second order derivatives $\bar E_{N}^{(i)}(z)$ $(i=1,2)$ at $N=50$ in the plane $z=t^{3/2}\in \mathbb{C}$.  The blue circles indicate zeros of $\bar E_{N}^{(1)}(z)$ and red filled circles zeros of $\bar E_{N}^{(2)}(z)$.  We take zero of $\bar E_{N}^{(2)}(z)$ indicated by the arrow as the best optimal solution.  This solution corresponds to the filled circle shown in FIG. 7(b).}
\end{figure}
As the order increases, the numbers of stationary points in each sets increase and the intersection area becomes a dense set of zeros, which we call the center of zeros.   The function is smoothest there and the amplitude is smallest.  The two points indicated in FIG. 7(a),(b) are located at this intersection area.  The red filled circle indicated by the arrow in FIG. 8 is the red filled circle plotted in FIG. 7(b).

These observations help us handling PMS in the complicated proliferation of stationary or almost stationary points.   To summarize, pick out the point in the center of zeros satisfying either (i) $\bar E^{(1)}_{N}(t)=0$ with $|\bar E^{(2)}_{N}(t)|\ll 1$ or (ii) $\bar E^{(2)}_{N}(t)=0$ with $|\bar E^{(1)}_{N}(t)|\ll 1$.   This prescription may be regarded as a variant of the PMS criterion and we continue using the term PMS in what follows.  
Under the above criteria, we have estimated ${\cal E}$ to $300$th orders.  The result of the estimation is plotted in FIG. 9 where the vertical axis labels $\log_{10}|E^{*}-{\cal E}|$.  
\begin{figure}
\centering
\includegraphics[scale=0.85]{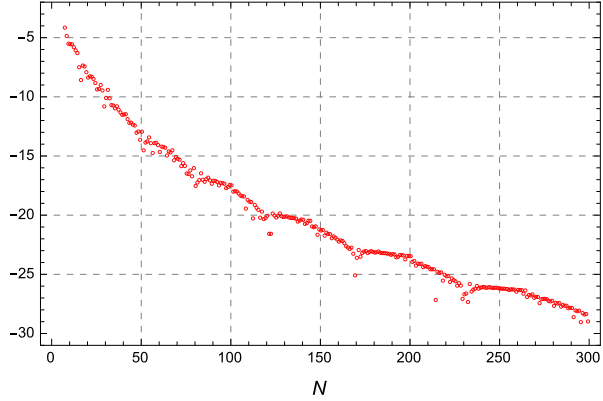}
\caption{Estimates of ${\cal E}$ up to $300$th orders.  The vertical axis indicates $\log_{10}|E^{*}-{\cal E}|$ where $E^{*}$ means the estimate.}
\end{figure}

We observe the expected growth of the accuracy with the orders.  Due to the oscillation property of $\bar E_{N}(t)$, there is a periodic pattern and the length of the period becomes longer as the order increases.  In the same time, the rate of accuracy growing becomes gradually slow down, though there seems to be no limit of approaching to ${\cal E}$.

The effective region of $\bar E_{N}(t)$ shrinks as the order increases.  This is already seen in FIGs. 2, 3 and 4.  Accordingly, the estimation point moves to the origin with the order.  The value of the estimation point $t^{*}$ is plotted in FIG. 10.  The precise fitting of the data is not allowed since the distribution of data is somewhat complicated with periodic structure.  We just remark that, from estimates for $N=270\sim 300$ where the data are rather steady, $t^{*}$ tends to zero roughly like $\sim N^{-0.56}$.
\begin{figure}
\centering
\includegraphics[scale=0.85]{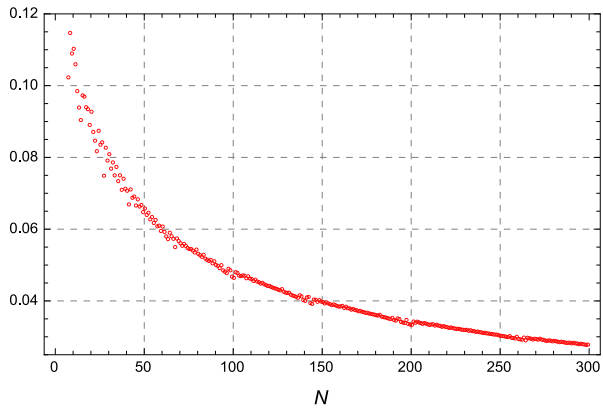}
\caption{Plots of the value of $t=t^{*}$ at which ${\cal E}$ is estimated.}
\end{figure}

\subsection{Comparison with the linear delta expansion}
In this subsection, we compare the result obtained in the generalized binomial transform with the one obtained in the linear delta-expansion.  The computation of the ground state energy in linear delta-expansion has been already done by Janke and Kleinert up to $251$th order  \cite{jk} (The expansion technique is called the variational perturbation theory).  Let us first explain the work with focusing on the related part.

The PMS criterion works more straightforwardly in the linear delta-expansion.  This is understood by  plotting the function $E_{LDE, N}(m)$ (see (\ref{linear delta expansion})).  Omitting the graph plots, we note that the best estimation point given as the stationary point or the inflection point is always the one {\it at the largest value of $\lambda/m^3$}, since the oscillation amplitude becomes smallest there.   It is interesting to see the distribution of zeros in the complex extension of the first order derivative $E_{LDE,N}^{(1)}(z)=[(\partial/\partial \log m^{-2})E_{LDE,N}]|_{\lambda/m^3\to z}$ where $z=\lambda/m^3$.
\begin{figure}
\centering
\includegraphics[scale=0.75]{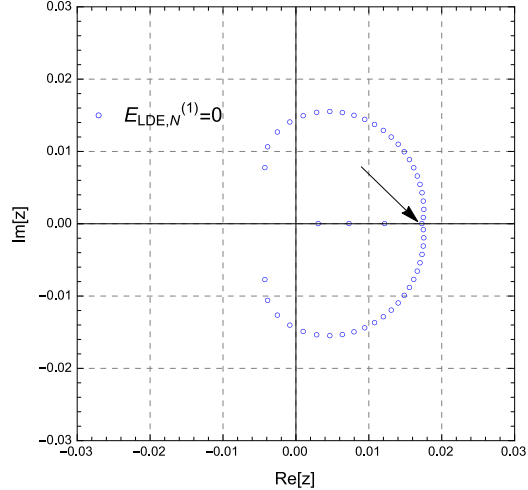}
\caption{The plots of zeros of the first derivative $\bar E_{LDE, N}^{(1)}(z)$ at $N=50$.  Here the argument $z$ is the complex extension of $\lambda/m^3$.}
\end{figure}
From the plot shown in FIG. 11, we find that the intersection of the zero point set on the positive real axis and the set extending in arc-form on the right half plane occurs at the largest real zero (The point indicated by the arrow in FIG. 11).  Thus, also in the linear delta expansion, the estimation point lies on the intersection of the two sets.

\begin{table*}
\caption{Estimation of ${\cal E}=0.667986259155777108270962016919860199430404936\cdots$ in sample orders in linear delta expansion and generalized binomial transform approaches.  The exact figures in each digits are written in Roman style while figures in the last two digits including errors are written in Slanted style.}
\begin{center}
\begin{tabular}{cll}
\hline\noalign{\smallskip}
$N$  &\quad linear delta expansion &\quad generalized binomial transform   \\
\noalign{\smallskip}\hline\noalign{\smallskip}
$10$   &\quad  0.66798{\it 57}  &\quad 0.66798{\it 37}  \\
$15$   &\quad  0.667986{\it 30} &\quad 0.66798{\it 58}  \\
$20$   &\quad  0.6679862{\it 62} &\quad 0.6679862{\it 68}  \\
$25$   &\quad  0.667986259{\it 20} &\quad 0.66798625{\it 79}  \\
$50$   &\quad  0.667986259155{\it 92}  &\quad 0.6679862591557{\it 58}   \\
$100$   &\quad  0.667986259155777{\it 05}  &\quad 0.66798625915577710{\it 53}  \\
$150$   &\quad  0.667986259155777108{\it 39}  &\quad 0.66798625915577710827{\it 14}  \\
$200$   &\quad  0.667986259155777108270{\it 34}  &\quad 0.6679862591557771082709{\it 59}  \\
$250$   &\quad  0.6679862591557771082709{\it 57}  &\quad 0.6679862591557771082709620{\it 22}   \\
$300$   &\quad  0.667986259155777108270962{\it 48}  &\quad 0.6679862591557771082709620169{\it 28}  \\
\noalign{\smallskip}\hline
\end{tabular}
\end{center}
\end{table*}
The estimation result at the largest stationary point is plotted in FIG. 12 and the numerical results in both schemes (linear delta and binomial) are tabulated in table 3.  In ref. \cite{jk}, the highest order studied is $251$th and the result is quoted as ${\cal E}=0.66798625915577710827096$, which is confirmed by us with the explained protocol.

In FIG. 12, we have also plotted the results in generalized binomial transform for the sake of the comparison.  At low orders up to, say roughly $20$th, the result from linear delta-expansion is slightly more accurate.  Then, as the order increases, the crossover occurs and at large orders, the results from binomial transform become superior.  Since the sequence in binomial scheme achieves higher accuracy than the sequence (which convergence is proved) from linear delta expansion, its convergence is now verified.  
\begin{figure}
\centering
\includegraphics[scale=0.85]{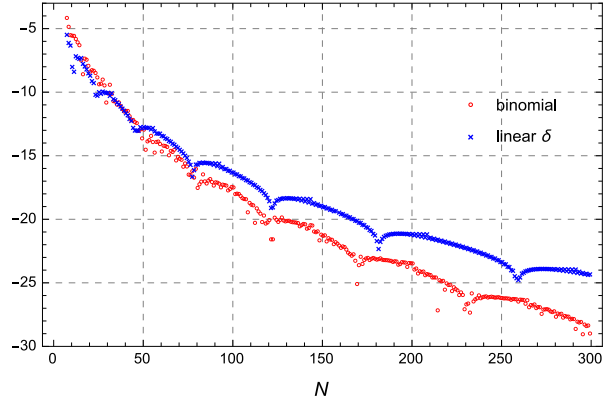}
\caption{The plots of $\log_{10}|E^{*}-{\cal E}|$ in the linear delta-expansion and binomial transform.}
\end{figure}

In the case of the anharmonic oscillator, we have used no technique special to the model.  The high accuracy of the estimation comes from the analytic structure with respect to $m^2$ of the strong coupling series of the ground state energy \cite{tsuika},
\begin{equation}
E(m,\lambda)={\cal E}\lambda^{1/3}\{1+e_{1}(m^2 \lambda^{-2/3})+e_{2}(m^2\lambda^{-2/3})^2+\cdots\}.
\label{str}
\end{equation}  
Linear delta expansion eliminates lower order terms since $\{m^2(1-\delta)\}^{L}=0$ for $N\ge L$.  Also in binomial method, ${\cal B}_{N}[(m^2)^{L}]={N \choose -L}t^L=0$ since ${ N \choose -L}=\Gamma(N)/\{\Gamma(-L+1)\Gamma(N+L+1)\}=0$ for $L=1,2,3,\cdots$.  That is, both of the linear delta expansion and binomial transform methods receive an advantage from the fact that $(m^2)^{n}\to 0$ $(n=1,2,3,\cdots)$ after the expansion or transformation.  This is the reason why linear delta expansion and binomial transform with respect to $m^2\lambda^{-2/3}$ yield accurate estimates unlike the case of Laplace integral where corrections of power series in $1/t$ remain. 

\subsection{Estimation of the strong coupling coefficients}
Analyticity with respect to $m^2$ expressed in (\ref{str}) can be numerically confirmed by binomial transform, as we can see below:  Assume that there are fractional power-like terms and let the leading one be $const\times (m^2/\lambda^{2/3})^{\Delta}$ ($\Delta>0$).   Then
\begin{equation}
{\cal B}[E(m,\lambda)]={\cal E}\lambda^{1/3}\{1+const\times t^{-\Delta}+\cdots\}.
\label{binomial-str}
\end{equation}  
The leading correction from $t^{-\Delta}$ must then be observed in $\bar E_{N}(t,\lambda)$ if it would exist, as would be seen in the plot of $\psi_{0}$ in FIG. 1 where $t^{-1}$ correction is active.  But the numerical plot shown in FIG. 6 does not imply any power like correction.  This means that the terms of fractional powers are absent in the strong coupling expansion.  Thus, the expansion (\ref{str}) is ensured even in our numerical study.

The coefficient $\alpha_{k}={\cal E}e_{k}$ of the series (\ref{str}) can be estimated in the following way:  As the first example, we illustrate the estimation of $\alpha_{1}$.  Setting $\lambda=1$, consider the derivative of $E(m^2,1)$ with respect to $m^2$ denoted as $ \partial E(m^2,1)/\partial m^2:=E^{'}(m^2,1)$.  We obtain at small $m^2$
\begin{equation}
E^{'}(m,1)=\alpha_{1}+2\alpha_{2}m^2+3\alpha_{3}(m^2)^2\cdots,
\end{equation}  
and at large $m^2$
\begin{equation}
E_{N}^{'}(m,1)=\sum_{n=0}^{N}a_{n}(-\frac{3n-1}{2})(1/m^2)^{(3n+1)/2}.
\end{equation} 
Then, the binomial transform eliminates corrections of integer powers of $m^2$ in (\ref{str}) and may simply leave 
\begin{equation}
{\cal B}[E^{'}(m,1)]\sim \alpha_{1},
\label{alpha1}
\end{equation} 
at a certain region where (\ref{alpha1}) is expected to be recovered by ${\cal B}[E^{'}_{N}(m,1)]$.  
As in the same manner of estimating ${\cal E}$, we have carried out estimation of $\alpha_{1}$ by substituting
\begin{equation}
{\cal B}[E_{N}^{'}(m,1)]=\sum_{n=0}^{N}a_{n}(-\frac{3n-1}{2}){ N \choose \frac{3n+1}{2}}t^{(3n+1)/2}
\end{equation}
into ${\cal B}[E^{'}(m,1)]$ and using PMS to pick out the optimal solution for $\alpha_{1}$.  For higher order coefficients, using the derivatives of $E(m^2,1)$ with respective $m^2$, we can estimate $\alpha_{2}$, $\alpha_{3}$ and so on.  We tabulate the results in table 4.  
\begin{table}
\caption{Estimation results of coefficients $\alpha_{k}={\cal E} e_{k}$ $(k=1,2,3,4,5)$ of the strong coupling expansion of the ground state energy at $250$th and $300$th orders.  The results are expressed to the digit of order $10^{-28}$ the same order of correct ${\cal E}$ estimated at $N=300$ (cf. table 3).}
\begin{center}
\begin{tabular}{cr}
\noalign{\smallskip}\hline\noalign{\smallskip}
$\alpha_{1}\,(N=250)$   &\quad   0.1436687833808649100203190808  \\
$\alpha_{2}\,(N=250)$   &\quad  -0.0086275656808022791279635744 \\
$\alpha_{3}\,(N=250)$   &\quad   0.0008182089057563495424151582 \\
$\alpha_{4}\,(N=250)$   &\quad  -0.0000824292171300772199109668 \\
$\alpha_{5}\,(N=250)$   &\quad   0.0000080694942350409647544789 \\
\noalign{\smallskip}\hline\noalign{\smallskip}
$\alpha_{1}\,(N=300)$   &\quad   0.1436687833808649100203191272  \\
$\alpha_{2}\,(N=300)$   &\quad  -0.0086275656808022791279637461 \\
$\alpha_{3}\,(N=300)$   &\quad   0.0008182089057563495424155947 \\
$\alpha_{4}\,(N=300)$   &\quad  -0.0000824292171300772199118949 \\
$\alpha_{5}\,(N=300)$   &\quad   0.0000080694942350409647560181 \\
\noalign{\smallskip}\hline
\end{tabular}
\end{center}
\end{table}
Having compared our results at $N=250$ and $300$, we consider that the figures at $N=250$  to $10^{-24}$ order are correct for $\alpha_{k}$ ($k=1,2,3,4,5$).  

Let us compare our results with those obtained by Janke and Kleinert at order $N=251$ \cite{jk}.  Their results are
\begin{eqnarray}
\alpha_{1}^{JK}&=&0.1436687833808649100203,\nonumber\\
\alpha_{2}^{JK}&=&-0.008627565680802279128,\nonumber\\
\alpha_{3}^{JK}&=&0.000818208905756349543,\nonumber\\
\alpha_{4}^{JK}&=&-0.000082429217130077221,\nonumber\\
\alpha_{5}^{JK}&=&0.000008069494235040966.
\end{eqnarray}
As in the case of ${\cal E}$, our results for $\alpha_{k}$ $(k=1,2,3,4,5)$ are more accurate than $\alpha_{k}^{JK}$ about by $2\sim 3$ digits.  We thus conclude that, as long as the order is high, the estimate of strong coupling coefficients is better in binomial transform.

\subsection{Binomial transform with respect to $\lambda$}
The computation of the ground state energy has so far been done by taking the energy as a function of $m^2$.  
We investigate here the transform with respect to the coupling constant by taking the energy as a function of $\lambda$.  In this point of view, the corrections have fractional powers in the strong coupling region as seen from (\ref{str}).  Hence, we can test the flexibility of the generalized binomial transform approach by the application to such a complicated case.  To dilate the region around the strong coupling limit $\lambda=\infty$, we re-scale $\lambda=g/(1-\delta)$ and expand the energy function in $\delta$ to the relevant order of perturbation series.

For the sake of notational simplicity, we set $m^2=1$.  Then, the behavior of $E(1,\lambda)$  at $\lambda\gg 1$ reads from (\ref{str})
\begin{equation}
E(1,\lambda)={\cal E}\lambda^{1/3}(1+e_{1}\lambda^{-2/3}+e_{2}\lambda^{-4/3}+\cdots),
\label{str2}
\end{equation}  
and at $\lambda\ll 1$
\begin{equation}
E_{N}(1,\lambda)=a_{0}+a_{1}\lambda+a_{2}\lambda^{2}+\cdots+a_{N}\lambda^{N}.
\end{equation} 
We investigate the computation of ${\cal E}$ in most part without using the values of the exponents of the corrections to the asymptotic term.  We thus start with
\begin{equation}
E(1,\lambda)={\cal E}(\lambda^{1/3}+e_{1}\lambda^{-\theta_{1}}+e_{2}\lambda^{-\theta_{2}}+\cdots).
\label{str3}
\end{equation}  

The basic result needed for the transform is then
\begin{equation}
{\cal B}[\lambda^{s}]={ N \choose s}g^s.
\end{equation}
Although the transform improves the simulation task of $\bar E(1,g)$ via $\bar E_{N}(1,g)$, to achieve a good accuracy, we need to reduce the correction.  So we must somehow estimate first the values of the exponents $\theta_{i}$ up to, say, a first few $i$.  It has turned out, though, that higher order $\theta_{i}$ is difficult to estimate precisely in stable and systematic manner.  Here we suffice ourselves with the reduction of the first order correction by the estimation of $\theta_{1}=1/3$ and make use of the result for the estimation of ${\cal E}$.  

We start with noting that, since the leading order correction has the exponent $1/3$ known on dimensional grounds, the leading term can be eliminated in the following combination,
\begin{equation}
\bar E-3\bar E^{(1)}={\cal E}\Big[e_{1}{N \choose \theta_{1}}(1+3\theta_{1})g^{-\theta_{1}}+e_{3}{N \choose \theta_{2}}(1+3\theta_{3})g^{-\theta_{3}}+\cdots\Big],
\end{equation}
with $\bar E^{(1)}=(\partial/\partial \log g)\bar E(1,g)$ understood.  Here we have used the fact that ${\cal B}_{N}[\lambda^{-\theta_{2}}]={\cal B}_{N}[\lambda^{-1}] =0$.  Equally, all the terms with negative integer powers are eliminated by the binomial transform.  To avoid notational complexity, we re-parametrize exponents as $\theta_{1}(=1/3),\theta_{2}(=5/3),\theta_{3}(=7/3),\theta_{4}(=11/3)$ etc.  Then in general, it holds that
\begin{equation}
\prod_{i=0}^{L}\Big[1+\frac{1}{\theta_{i}}\frac{\partial}{\partial \log g}\Big]\bar E=const.\times g^{-\theta_{L+1}}+O(g^{-\theta_{L+2}}),
\label{reduction}
\end{equation}  
where $\theta_{0}=-1/3$.  Now, taking the following quotient and expanding it in $1/g$, we find
\begin{equation}
Q_{L}=\frac{\prod_{i=0}^{L}\Big[1+\frac{1}{\theta_{i}}\frac{\partial}{\partial \log g}\Big]E^{(1)}}{\prod_{i=0}^{L}\Big[1+\frac{1}{\theta_{i}}\frac{\partial}{\partial \log g}\Big]E}=-\theta_{L+1}+\cdots,
\end{equation} 
where the dots means the correction of order $O(g^{-\theta_{L+2}+\theta_{L+1}})$.  We here concern with the case $L=0$, giving at large $g$,
\begin{equation}
Q_{0}=\frac{\Big[1-3\frac{\partial}{\partial \log g}\Big]E^{(1)}}{\Big[1-3\frac{\partial}{\partial \log g}\Big]E}=-\theta_{1}+\cdots,
\label{quotient}
\end{equation} 
The quotient $Q_{0}$ is a divergent series in $g$ which we denote as $Q_{0,N}$, where $\bar E_{N}$ and $\bar E_{N}^{(1)}$ are used in the places of $\bar E$ and $\bar E^{(1)}$.   The function $Q_{0,N}$ thus defined looks like the Pad\'e-type rational function but it is actually not.  Hence by first expanding $Q_{0,N}=[1-3\frac{\partial}{\partial \log g}]E_{N}^{(1)}/[1-3\frac{\partial}{\partial \log g}]E_{N}$ in $g$, we construct diagonal Pad\'e approximants to circumvent the zero-convergence-radius difficulty.  We then take the $g\to \infty$ limit of $Q_{0,N}[N/2,N/2]$ as the estimate of $\theta_{1}$.  That is, from (\ref{quotient}), 
\begin{equation}
\lim_{g\to \infty}Q_{0,N}[N/2,N/2]=-\theta_{1}.
\label{theta-estimate}
\end{equation} 
The result is plotted in FIG. 13 with the label "1p(opt-Pade)".  
\begin{figure}
\centering
\includegraphics[scale=0.85]{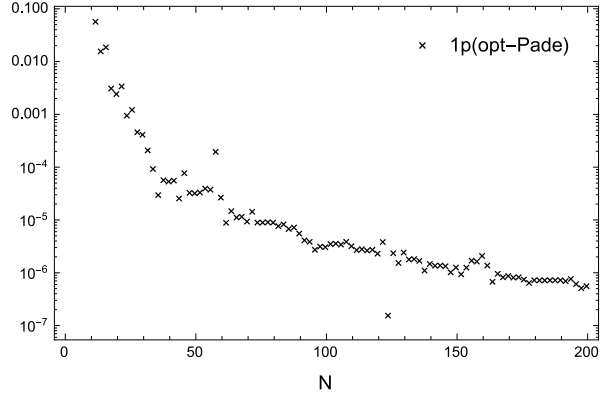}
\caption{The logarithmic plots of $|\theta_{1}-1/3|$ up to $200$th order.}
\end{figure}
The sequence does not exhibit clear shape and the convergence issue is not definitive.  However, the estimation of $\theta_{1}$ is accurate enough and we make use of the estimate at order $N$ to the estimate of ${\cal E}$ at the same order in the following manner.

One technical point to be payed attention is that the asymptotic of $\bar E$ in $g\to \infty$ is ${\cal E}{N \choose -1/3}g^{1/3}$ which is $g$ dependent.  The ground state energy may be estimated in this case by dividing $\bar E$ by ${N\choose -1/3}g^{1/3}$.  To improve the accuracy, however, we again use first order reduced function $[1+(1/\theta_{1})\partial/\partial \log g)]\bar E$.  We thus study the combination $(\bar E +(1/\theta_{1})\bar E^{(1)})/\{{N\choose -1/3}g^{1/3}\}$ which tends to ${\cal E}$ in the large $g$ limit,
\begin{equation}
\lim_{g\to \infty}\frac{\bar E+\frac{1}{\theta_{1}}\bar E^{(1)}}{{N\choose -1/3}g^{1/3}}={\cal E}.
\label{e-estimate}
\end{equation}

Now, in the use of $\bar E_{N}$ and $\bar E_{N}^{(1)}$, the above limit does not hold even when we employ Pad\'e approximants of any $[\rho/\tau]$ element in the numerator $\bar E_{N}+\frac{1}{\theta_{1}}\bar E_{N}^{(1)}$ of (\ref{e-estimate}).  This is reasonable, since beyond narrow effective region of the series expansion Pad\'e approximants take extrapolation affected by the  highest orders $g^{\rho}$ and $g^{\tau}$ of the numerator and the denominator, giving the behavior $(\bar E+\frac{1}{\theta_{1}}\bar E^{(1)})[\rho/\tau]\sim const.\times g^{\rho-\tau}$ which exponent cannot agree with $1/3$, the power of $g^{1/3}$.  The expected asymptotic behavior $\sim g^{1/3}$ should occur at a certain region where $g$ is not so large.  The reliable region is indicated by the plateau in the combination (\ref{e-estimate}) and the plateau serves an optimal estimation point of ${\cal E}$ under PMS.  Due to the smallness of the exponent $1/3$ which should be recovered by Pad\'e approximants, best choice is the diagonal one.  Then we find that the optimal point occurs at either the extremum or the reflection points with small enough gradient.  We also keep the estimate when poles on the positive real axis exist.  
\begin{figure}
\centering
\includegraphics[scale=0.85]{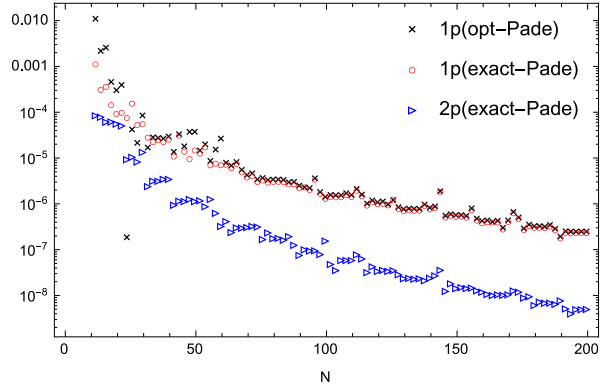}
\caption{The logarithmic plots of $|E^{*}-{\cal E}|$ up to $200$th order in the three methods:  The "1p(opt-Pade)" shows the result from the first order reduction via $[1+3(\partial/\partial\log g)]\bar E/\{{N\choose -1/3}g^{1/3}\}$ with the order dependent optimal $\theta_{1}$.  The "1p(exact-Pade)" sequence shows the result via $[1+3(\partial/\partial\log g)]\bar E/\{{N\choose -1/3}g^{1/3}\}$ with exact $\theta_{1}$ and the "2p(exact-Pade)" via $[1+3/5(\partial/\partial\log g)][1+3(\partial/\partial\log g)]\bar E/\{{N\choose -1/3}g^{1/3}\}$ with exact $\theta_{1}$ and $\theta_{2}$.}
\end{figure}

The result is plotted  in FIG. 14 by the black cross labeled by "1p(opt-Pade)".  In FIG. 14, we have also plotted, for the comparison with the idealistic same optimization procedure, the result at the first and second order reductions of corrections with exact exponents.  The "1p(exact-Pade)" labelled sequence shows the result via the quotient $[1+3(\partial/\partial\log g)]\bar E/\{{N\choose -1/3}g^{1/3}\}$ and the "2p(exact-Pade)" via the quotient $[1+3/5(\partial/\partial\log g)][1+3(\partial/\partial\log g)]\bar E/\{{N\choose -1/3}g^{1/3}\}$.  We now see that the optimal solution of $\theta_{1}$ provides almost same accuracy with the use of the exact value of $\theta_{1}$.  The reduced function exact to the second order brought more accuracy as it should.  

From the plots, all the above three sequences to $N=200$ appear to converge from the plots.  
One evidence of the convergence of the sequences comes from the behaviors of estimation point $g=g^{*}$.  All three sequences of $g^{*}$ show gradual increase with the order.  This means that the approximate region is indeed extrapolated to larger $g$ and, consequently, the convergence becomes quite conceivable due to the disappearance of the binomial coefficient of $g^{-\theta_{i}}$ $(i=1,2,3,\cdots)$, ${N \choose -\theta_{i}}$, in the $N\to \infty$ limit.

The reduction of the correction provided by (\ref{reduction}) gives further accurate estimates as the order of reduction $L$ is increased.  By using exact values of $\theta_{i}$ to more many orders, we obtain
\begin{eqnarray}
L=10&:& 0.66798625915577710858725991,\\
L=20&:& 0.66798625915577710827111996,\\
L=30 &:& 0.66798625915577710827096268,
\end{eqnarray}
at $N=250$.  The results are respectively exact to $10^{-18}$, $10^{-20}$ and $10^{-24}$ orders.  The last result achieved the accuracy with the almost same order with the one in linear delta expansion.  The case $L=40$ gives less accurate result at $N=250$ than the case $L=30$.  However, we confirmed that, at higher order $N=300$, the case $L=40$ exceeds in the accuracy than the case $L=30$: ${\rm error}=1.8\times 10^{-26}\, (L=30)$ and ${\rm error}=9.6\times 10^{-27}\, (L=40)$.  Thus, increasing the incorporated exponents indeed improves the accuracy while the effect manifests at larger orders.

\section{Concluding remarks}
We have explored the generalized binomial transform in the application to a Laplace integral and the quantum anharmonic oscillator in the perturbative framework.  Let us summarize the investigation below.

In the Laplace integral function (\ref{Lap}), binomial transform eliminated the logarithmic singularity at $M=0$ and left the series in $1/t$.  This is an advantage of the transformed series.  The limit $\lim_{M\to 0+}f(M)$ is approximately computed from binomial transform of $f(M)$ expanded in $1/M$ valid at $M\gg 1$.   Here the use of the technique to subtract corrections in assumed power series was critical to achieve a good accuracy.  The extrapolation to the infinite order by a simple fitting ansatz predicts a slightly different value, which is caused by the non-convergence nature of divergent series.  To go beyond that difficulty, we found that diagonal Pad\'e approximants of the transformed series provided excellent estimation at finite orders and solid evidence of convergence to the exact limit.  

In the case of the anharmonic oscillator, the ground state energy computation is successful in the naive transformed function by detecting the optimal stationary or almost stationary points among many candidates under PMS.  The presence of a few to many stationary points is not a serious drawback of binomial transform or linear delta expansion, since the optimal and best estimation comes from the point with the smallest amplitude and is thereby detectable.   The accuracy of estimate is periodically improving as the order grows and the convergence of the sequence is confirmed by the comparison with that in the linear delta expansion.  Also the coefficients in the strong coupling expansion could be estimated with the same level of accuracy.  The high level of accuracy has its origin to the fact that the energy function $E(m^2,\lambda)$ is analytic around $m^2=0$ \cite{tsuika}.  In this case, the subtractive-reduction of the corrections with PMS brings no essential improvement.

As the third study, from the point of view that the ground state energy is considered as a function of the coupling constant, we investigated  the binomial transform approach to the computation of ${\cal E}$.  Explicit reduction under the circumstance that $\theta_{i}$ is not known was carried out in the case of one-parameter ($\theta_{1}$) reduction.  Although the estimates became less precise compared with the previous two cases, obtained estimations were satisfactory.  It has been explicitly confirmed that the subtractive-reduction to higher order corrections is effective to obtain accurate estimates.  The problem is, of course, the precise values of the exponent $\theta_{i}$ is necessary.  If there is no confluent singularity, it is expected that accurate results would be obtained with the aide of approximate elimination of higher order corrections.  However, in models where a  new singularity represented by the correction to the scaling exponent exists, the estimation with high accuracy may become a hard task, in particular for divergent series.

In these three estimation tasks, we found that, when the transformed function has power like corrections in the target region of the argument (the region where $M\ll 1$ in the present work), non-convergent nature of divergent series available in the opposite accessible region does not allow the exact convergence of the sequence of estimates from transformed polynomial, though the improvement by the reduction of the correction produces accurate results.  The problem has been resolved by the Pad\'e approximant method by which the successful extrapolation beyond original narrow effective region was achieved.  Note that, in the study of the anharmonic oscillator based on the binomial transform with respect to $\lambda$, reduction of the first order correction was crucial for the estimation by the Pad\'e approximants.  In various physical models, the combined use of the correction-reduction and Pad\'e approximants may become useful tools for dealing with the divergent series of a given function, usually accessible in the perturbative side where $M\gg 1$, for the quantitative computation of the function in the target region where $M\ll 1$.



\begin{thebibliography}{99}

\bibitem{yam} H. Yamada, Phys. Rev. D76, 045007 (2007).
\bibitem{yam2} H. Yamada, Phys. Rev. E90, 032139 (2014).
\bibitem{yam3} H. Yamada, Braz J. Phys. 45, 584 (2015).
\bibitem{knp} J-L. Kneur, A. Neveu and M. B. Pinto, Phys. Rev. A69, 053624 (2004), and the references therein.
\bibitem{sz} R. Seznec and J. Zinn-Justin, J. Math. Phys. 20, 1398 (1979); J. Zinn-Justin, Appl. Num. Math. 60, 1454 (2010) (arXiv:1001.0675 [math-ph]).
\bibitem{lattice} A. Duncan and M. Moshe, Phys. Lett. 215B, 352 (1988); A. Duncan and H. F. Jones, Nucl.Phys. B320,189 (1989); I. Buckley and H. F. Jones, Phys. Rev. D 45, 654 (1992); I. Buckley and H. F. Jones, Phys. Rev. D 45, 2073 (1992); J. Akeyo and H. F. Jones, Phys. Rev. D 47, 1668 (1993).
\bibitem{stev} P. M. Stevenson, Phys. Rev. D23, 2916 (1981); Nucl. Phys. B203, 472 (1982).
\bibitem{baker} G. A. Jr. Baker and P. R. Graves-Morris, {\it Pad\'e approximants}, 2nd. edition, Cambridge Unversity Press, (1996).
\bibitem{bw} C.M. Bender and T.T. Wu, Phys. Rev. 184, 1231 (1969); Phys. Rev. D7, 1620 (1973).
\bibitem{bgn} B. Bellet, P. Garcia and A. Neveu, Int. J. Mod. Phys. A11, 5587 (1996).
\bibitem{gks} R. Guida, K, Konishi and H. Suzuki, Ann. Phys. 241, 152 (1995); Ann. Phys. 249, 109 (1996).
\bibitem{vin} F. Vinette and J. Cizek, J. Math. Phys. 32, 3392 (1991).
\bibitem{wen} Weniger also proposed a method admitting accurate computation of the ground state energy of the quartic, sextic and octic anharmonic oscillator.  See E. J. Weniger, Phys. Rev. Lett. 77, 2859 (1996).
\bibitem{neveu} A. Neveu, Nucl. Phys. (Proc. Suppl.) B18, 242 (1990).
\bibitem{comment1} However, this generalization complicates the concept of interpolation.  The  Hamiltonian becomes
\begin{eqnarray*}
H(\delta;a,b_{n})&=&\frac{1}{2}\big(\frac{\partial \phi}{\partial q}\big)^2+\frac{m^2}{2}(1-\delta)\{1+(a-1)\delta\\
& &+\sum_{n=1}b_{n}\delta^{n+1}\}\phi^2+\delta\lambda \phi^4,
\end{eqnarray*}
and the physical interpretation is obscured.
\bibitem{jk} W. Janke and H. Kleinert, Phys.Rev. Lett. 75, 2787 (1995).
\bibitem{tsuika} B. Simon, Ann. Phys. (N.Y.) 58, 76 (1970).
\end{thebibliography}
\end{document}